\documentstyle[12pt,epsf,epsfig]{article}

\newcommand{\beq}{\begin{equation}}
\newcommand{\eeq}{\end{equation}}

\begin{document}

\begin{center}
{\large\bf A HERETICAL VIEW\\

\vskip 0.5true cm
 ON LINEAR REGGE TRAJECTORIES}

\vskip 0.6true cm
{\large\bf Dmitri Diakonov$^{*\,\diamond}$ and Victor Petrov$^*$}\\

\vskip 0.5true cm
$^*$ Petersburg Nuclear Physics Institute, Gatchina 188300 Russia\\
\vskip 0.2true cm
$^\diamond$ NORDITA, Copenhagen, DK-2100 Denmark 
\end{center}

\vskip 0.4true cm
\begin{abstract}
We discuss  a possibility that linear Regge trajectories originate
not from gluonic strings connecting quarks, as it is usually assumed,
but from pion excitations of light hadrons. From this point of view, 
at large angular momenta both baryons and mesons lying on linear
Regge trajectories are slowly rotating thick strings of pion field,
giving rise to a universal slope computable from the pion decay 
constant. The finite resonance widths are mainly due to the semiclassical 
radiation of pion fields by the rotating elongated chiral solitons. 
Quantum fluctuations about the soliton determine a string theory which, 
being quantized, gives the quantum numbers for Regge trajectories.\\

This is a reprint, with a minimal editing, of the paper entitled ``Rotating 
chiral solitons lie on linear Regge Trajectories", which appeared as a 
preprint LNPI-88-1394 (May 1988) but was not published.
\end{abstract}

\section{Introduction}

A long-standing problem of strong interactions is to understand
linear Regge trajectories. Experimentally, most hadrons containing
light quarks with spin $J$ have partners with the same quantum
numbers but spin $J+2$, $J+4,\ldots$~. Lines connecting these
partners in the Chew--Frautchi plot have, with a few exceptions,
parallel slopes $\alpha'\simeq 0.8-0.9\,{\rm GeV}^{-2}$.  Certain
trajectories seem to be parity and/or signature degenerate. In
some cases parallel ``daughter" trajectories are clearly seen. The
present experimental status is shown in Figs. 1--4 where we plot
all known non-strange hadrons from the 1986 Review of Particle
Properties.

The usual qualitative explanation of resonances lying on linear
trajectories is that they are rotating confining gluonic strings or
flux tubes attached to quarks at the end points moving with the
speed of light \cite{1,2}. The idea is rather old and is usually taken
for granted. Good or bad, it ignores the spontaneous
chiral symmetry breaking in QCD. Owing to it,
nearly massless ``current" quarks get a dynamical mass around
$350\,{\rm MeV}$, and the lightest hadrons are (pseudo) Goldstone
pions.  It costs very little energy to produce a pion and hence the 
would-be strings of the pure glue world have to break \cite{3}. 

Even if one takes for granted the flux tube model for
Regge trajectories,  there is still a long way to the realistic  
spectrum. Much work in that direction has been done \cite{2,6,7,8,9} 
but to our knowledge many important questions remain unanswered.
Why would $350\,{\rm MeV}$ quarks bound by a flux tube
with a $\sqrt{\sigma}=1/\sqrt{2\pi\alpha'}\simeq 425\,{\rm MeV}$
string tension move with the speed of light is not very clear, but if
they do not, the trajectories are not linear. Where, geometrically, 
is the baryons' third quark? Is it close to one of the ends or smeared 
over the whole string? Why some trajectories are degenerate in 
signature and/or parity and some are not? 

To point out one difficulty of the flux tube model, we
note that the Regge spectrum apparently ``knows" about the
spontaneous chiral symmetry breaking. For instance, the pion trajectory 
starts from zero intercept exhibiting the Goldstone nature of pions.
Another example: the baryon $1/2^+$ and $1/2^-$ trajectories seem
to be almost degenerate, except for their first members, $N(940,1/2^+)$
and $N(1535,1/2^-)$. These particles cannot be degenerate if the
chiral symmetry is spontaneously broken. Therefore, the
$1/2^{\pm}$ trajectories have to split at the end. However the
gluonic string would ``feel" the chiral symmetry breaking
only through quark loops, which is a $O(N_f/N_c)$ effect \cite{10}.
This quantity is considered to be small since it determines the
ratio of the widths to the masses of resonances. But the ugly
departure of the $1/2^-$ trajectory from a straight line does not
look like a small effect. 

In this paper we present an alternative view on linear Regge
trajectories as due to the rotating pion fields. Let us give
a few qualitative arguments in favor of such a heretical idea.

First, let us consider a ``soft" (as opposed to rigid) rotating
body (a hadron) and try to minimize its energy for given angular 
momentum $J$. Evidently, the minimum corresponds to a situation when 
the lightest piece of a body rotates at large distances around the heavy 
remnant. In other words, if a high $J$ hadron is considered 
as an excitation of a low $J$ one, the easiest way to get it is to excite 
the lightest degree of freedom, that is the pion field.

Second, large spin hadrons have presumably large sizes (at least
in one direction) owing to the centrifugal forces. It means that
fields inside such a hadron are slowly varying, in other words 
they have low momenta. The only degree of freedom of the 
strong interactions that  survives at low momenta is the 
pseudo-Goldstone pion field. Therefore, it seems natural to describe 
the high $J$ hadrons in terms of pion excitations.

Third, it is known that even the lowest member of the nucleon
Regge trajectory, {\it i.e.} the nucleon itself, can be understood as a
soliton of the pion field \footnote{The Skyrme model is the simplest
realization of this idea~\cite{11}. A more fine model of a nucleon
soliton has been proposed recently which seems to be more
satisfactory, both philosophically and quantitatively~\cite{12}.}. 
If the nucleon is a static chiral soliton, one is led to consider
high $J$ excitations of nucleons as rotating chiral solitons. 

Fourth, the pion itself belongs to a Regge trajectory which has
approximately the same universal slope. It should not be
accidental.

To complete the qualitative arguments, let us make a very rough
estimate of the mass of a rotating chiral soliton. Since the
characteristic size $r_0$ of the soliton will be shown to be
parametrically large in $J$, it is sufficient to use only the
 kinetic energy term of the effective chiral action,
\begin{eqnarray}
&& S_{\rm kin}\ =\ \frac{F^2_\pi}4 \int d^4x\sqrt{-g}\ g^{\mu\nu}
\mbox{ Tr }L_\mu L_\nu\ , \nonumber\\
&& L_\mu\ =\ iU^\dagger\partial_\mu U\ ,\quad U\ =\exp(i\pi^A\tau^A)\
, \quad F_\pi\simeq 93\mbox{ MeV }.
\end{eqnarray}
The energy of a static pion field configuration, as seen from
Eq.(1), grows linearly with its size $r_0$, $E_{\rm rest}\sim
F^2_\pi r_0$ (we ignore numerical coefficients). If a soliton
rotates, its rotation energy is $E_{\rm rot}\simeq J^2/2I$, where
$I$ is the moment of inertia. According to Eq.(1) $I$ grows as
the third power of the size, $I\sim F^2_\pi r^3_0$. The total
energy, $E_{\rm rest}+E_{\rm rot}$, is
$$ E\ \sim\ F^2_\pi r_0 +\frac{J^2}{F^2_\pi r^3_0}\ . $$
This function has a minimum at $r_0\sim\sqrt J /F_\pi$ (justifying
that $r_0$ is large at large $J$) with the value at the minimum
$E\sim F_\pi\sqrt J$. Hence, the mass $M_J$ of a rotating
soliton with a large angular momentum $J$ satisfies the equation
$$ J\simeq\alpha' M^2_J\ , \qquad \alpha'\sim\ F^{-2}_\pi\ . $$

Thus we obtain linear Regge trajectories from a simple
dimensional analysis. The Goldstone nature of pions is essential
in this derivation.

However simple, this derivation has loopholes. First, we have
introduced only one overall size $r_0$. It implies that the
rotating soliton is spherically-symmetric (as in the case of a
nucleon). But rotating hedgehogs have the same spin as isospin
\cite{11}. If one is interested in Regge trajectories with fixed
isospin one has to consider non-spherically--symmetric solitons.
Moreover, it is clear that centrifugal forces must stretch the
rotating soliton into something cigar-like. Therefore, the above
derivation has to be modified. This is performed in Sec.2. In
fact we arrive to an effective string theory for large $J$
chiral solitons with a calculable slope $\alpha'$.

Second, strictly speaking, there exists no classical solution of
the equations of motion, corresponding to a  stationary rotation.
Examples of this ``no go" theorem are presented in
Refs. \cite{13,14}. Its general cause is the same as in
electrodynamics: accelerated charges must radiate e.m. fields. In
our case a rotating chiral soliton, being an accelerated source of
isospin, must radiate pion fields. Therefore, one can speak only
of an approximately stationary rotation -- as far as the energy
loss per period is much less than the energy itself. However,
high $J$ Regge excitations are not expected to be stable.
Moreover, their lifetime can be calculated from the classical pion
radiation theory. This is performed in Sec. 3. A systematic way to
study quantum corrections to rotating solitons is outlined in
Sec. 4. Finally, Sec. 5 contains conclusions and an outlook.

\section{String-like chiral solitons}

Let us consider a chiral soliton rotating around the $z$ axis with
angular velocity $\omega$. We mark by primes coordinates in the
body-fixed frame. One has
\begin{eqnarray}
&& x'\ =\ x\cos\omega t+y\sin\omega t, \nonumber \\
&& y'\ =\ -x\sin\omega t+y\cos\omega t, \nonumber \\
&& z'\ =\ z, \nonumber \\
&& t'\ =\ t\ .
\end{eqnarray}
The metric tensor in the body-fixed frame is
\beq
g^{\alpha\beta}(x')=\frac{\partial x'^\alpha}{\partial
x^\mu}\frac{\partial x'^\beta}{\partial x^\nu}\ g^{(0)\mu\nu} =
\left(\begin{array}{cccr} 1 & \omega y' & -\omega x' & 0\\
\omega y' & \omega^2y'^2-1 & -\omega^2x'y' & 0\\
-\omega x' & -\omega^2x'y' & \omega^2x'^2-1 & 0\\
0 & 0 & 0 & -1 \end{array} \right). \eeq 
The kinetic energy term
of the effective chiral action (1) in this frame is 
\beq 
S= \frac{F^2_\pi}4 \int d^4x'g^{\mu\nu}(x')\mbox{ Tr } L_\mu
L_\nu\ , \qquad L_\mu=\ iU^\dagger \frac\partial{\partial x'^\mu}\ U\ .
\eeq
The Euler--Lagrange equation of motion reads
\beq
\frac\partial{\partial x'^\mu}\left[g^{\mu\nu}(x') U^\dagger(x')\
\frac\partial{\partial x'^\nu}\ U(x') \right]\ =\ 0\ . \eeq

We look for a solution of this equation, which is
(i)~time-independent in the body-fixed frame, (ii)~independent
of one coordinate (we choose it to be $x'$) orthogonal to the
rotation axis $z=z'$. With such restrictions Eq.(5) takes the
form
\beq
(1-\omega^2x'^2)\frac\partial{\partial y'}\left(U^\dagger
\frac\partial{\partial y'}U\right)+\frac\partial{\partial z'}
\left(U^\dagger\frac\partial{\partial z'}U\right)\ =\ 0\ .
\eeq
Introducing a Lorentz-contracted variable
\beq
\tilde y\ =\ \frac{y'}{\sqrt{1-\omega^2x'^2}}\ , \qquad \tilde
z\ =\ z'\ ,
\eeq
we rewrite Eq.(6) as
\beq
\frac\partial{\partial\tilde y}\left(U^\dagger\frac\partial{\partial
\tilde y}U\right)+\frac\partial{\partial\tilde z}\left(U^\dagger
\frac\partial{\partial\tilde z}U\right)\ =\ 0\ .
\eeq
Actually, it is the extremum condition for the transverse energy
or energy per unit length in the $x'$ direction,
\beq
E_\perp=\ \frac{F^2_\pi}4\int d\tilde yd\tilde z\mbox{ Tr}\left(
\partial_{\tilde y}U\partial_{\tilde y}U^\dagger +\partial_{\tilde z}U
\partial_{\tilde z}U^\dagger\right)\ .
\eeq

It has been shown in Ref. \cite{15} that Eq.(8) for the SU$_2$
chiral field has no non-trivial solutions other than embeddings
of two-dimensional grassmannian instantons. The simplest  
embedding is a two-dimensional hedgehog:
\begin{eqnarray}&& U\ =\ 
\exp(i\pi^A\tau^A)\ , \quad \pi^1=\ 0, \quad \pi^2=\frac{\tilde y}{\sigma} 
P(\sigma)\ ,\nonumber\\&& \pi^3\ =\ \frac{\tilde z}\sigma\ P(\sigma)\ , 
\quad \sigma = \sqrt{\tilde y^2+\tilde z^2}\ .
\end{eqnarray}
The transverse energy (9) becomes
\beq
E_\perp\ =\ \pi F^2_\pi\int\limits^\infty_0 d\sigma\
\sigma\left[\left(\frac{\partial P}{\partial\sigma}\right)^2
+\frac{\sin^2P}{\sigma^2}\right]\ .
\eeq

This functional has a non-trivial extremum corresponding to a
solution of a first-order ``self-duality" equation:
\beq
\frac{\partial P}{\partial\sigma}\ =\ -\frac{\sin P}\sigma\ .
\eeq
Its solution (with a unity two-dimensional topological charge)
is
\beq
P(\sigma)\ =\ 2\arctan\left(\frac{\sigma_0}\sigma\right)\ ,
\qquad P(0)\ =\ \pi\ ,
\eeq
where the transverse scale $\sigma_0$ is arbitrary. 
Eqs. (10,13) give a solution
of Eq.(8) as well. The integral in Eq.(9) is dimensionless,
therefore $E_\perp$ is independent of the string thickness
$\sigma_0$. We find at the extremum
\beq
E^{(0)}_\perp\ =\ 4\pi\,F^2_\pi\ .
\eeq
Substituting the ansatz (10) into Eq.(4) we get for the action
\beq
S\ =\ -E^{(0)}_\perp \int dt\int dx'\sqrt{1-\omega^2x'^2}\ .
\eeq
This expression coincides with the Nambu string action \cite{1}
\beq
S_{\rm Nambu}\ =\ -\frac1{2\pi\alpha'}\int d^2\xi \sqrt{(\dot
x_\mu x'_\mu)^2-\dot x^2_\mu x'^2_\mu}
\eeq
written for a rigid rotating string. Indeed, using the
parametrization
\beq
x_0\ =\ t\ =\ \xi_0\ , \qquad \dot x_ix'_i\ =\ 0\ ,
\eeq
(hence, $x_0=1,\ x'_0=0$) and taking into account that
$x^2_i=\omega^2s^2$ where $s$ is the length along the string
$(ds/d\xi_1=\sqrt{x'^2_i})$ and $\omega$ is the rotation angular
velocity, one gets from Eq.(16)
\beq
S_{\rm Nambu}\ =\ -\frac1{2\pi\alpha'}\int dt\int ds
\sqrt{1-\omega^2s^2}\ ,
\eeq
which coincides in form with Eq.(15), the Regge slope being
\beq
\alpha'\ =\ \frac1{2\pi E_\perp}\ =\ \frac1{8\pi^2F^2_\pi}\ .
\eeq

Let us check directly that the chiral soliton of the type given
by Eq.(10) leads to a linear Regge trajectory, without referring
to the Nambu lagrangian. To this end let us calculate the
angular momentum of the rotating pion field. The general 
expression is
\beq
J_i\ =\ \frac{F^2_\pi}2\int d^2x\ \varepsilon_{ijk}\mbox{ Tr }
\partial_0 U\,\partial_jU^\dagger x_k\ .
\eeq
Taking into account that the time dependence of the pion field
$U$ comes only through global rotation (see Eq.(2)) one can
rewrite $J_3$ in terms of body-fixed coordinates (denoted by a
prime):
\begin{eqnarray}
&& J_3\ =\ \frac{F^2_\pi\omega}2\int dx'dy'dz'\mbox{ Tr
}(x'L_{y'}-y'L_{x'})^2\ , \nonumber\\
&& L_{x',y'}\ =\ iU^+\ \frac\partial{\partial x',y'}\ U\ .
\end{eqnarray}
Using the ansatz (10) we find:
\begin{eqnarray}
J_3 &=& \frac{F^2_\pi\omega}2\int\frac{dx'x'^2}{\sqrt{1-\omega^2
x'^2}}\int d\tilde yd\tilde z\,\mbox{Tr }L^2_{\tilde
y}+O(\omega^0) \nonumber\\
&=& \omega E_\perp\int\frac{dx'x'^2}{\sqrt{1-\omega^2x'^2}}
+O(\omega^2)\ =\ \frac{\pi E_\perp}{2\omega^2}+O(\omega^0)\ .
\end{eqnarray}

We, thus, get the following relation between the angular
momentum and the angular velocity, which is typical for an
expandable string:
\beq
J\ =\ \frac{2\pi^2F^2_\pi}{\omega^2}\ .
\eeq

The larger $J$, the smaller is the rotation velocity $\omega$.
This is because the string length (along the $x'$ axis) is
$L=2/\omega$, its ends rotating with the speed of light.

At large $J$, $\omega$ is small, the chiral string is long, and
one can neglect departures from a simple infinite string-like
solution (10). It should be stressed that, strictly speaking,
Eq.(10) is not a solution of the full equation of motion (5). It
ceases to be a solution i)~at the end-points
$x'\approx\pm1/\omega$, ii)~far away from the string axis, {\it viz.}
at $y',z'\ge1/\omega$. However the corresponding corrections are
small in $\omega$ and are irrelevant to the calculation of the
Regge slope. Therefore, we neglect $L_{x'}$ in Eq.(21) and arrive
to Eq.(23).

Let us now calculate directly the energy of the rotating chiral
string. We have
\begin{eqnarray}
E &=& \frac{F^2_\pi}4\int d^3x\mbox{ Tr} \left(\partial_0U
\partial_0U^\dagger +\partial_iU\partial_iU^\dagger\right) 
\nonumber \\
&=& \frac{F^2_\pi}4\int dx'dy'dz'\mbox{ Tr}\left[(\omega^2x'^2
+1)\partial _{y'}U\partial_{y'}U^\dagger +\partial_{z'}U\partial_{z'}
U^\dagger\right] \nonumber\\
&=& \int dx'\sqrt{1-\omega^2x'^2}\ \frac{F^2_\pi}4\int d\tilde
yd\tilde z\mbox{ Tr}\left(\frac{1+\omega^2x'^2}{1-\omega^2x'^2}\
L^2_{\tilde y}+L^2_{\tilde z}\right) \nonumber\\
&=& E_\perp \int\frac{dx'}{\sqrt{1-\omega^2x'^2}}\ =\
\frac{\pi E_\perp}\omega\ =\ \frac{4\pi^2F^2_\pi}\omega\ .
\end{eqnarray}

Comparing this result with Eq.(23) we see that the mass squared
of a rotating chiral string grows linearly with the angular
momentum:
\beq
M^2_J\ =\ 8\pi^2F^2_\pi J\ =\ \frac1{\alpha'}J\ , \qquad \alpha'\
=\ \frac1{8\pi^2F^2_\pi}\ .
\eeq
This result coincides with that obtained previously
from comparison with the Nambu action and reveals what is called
a linear Regge trajectory.

Numerically we get from Eq.(25) $\alpha'\simeq1.45$ GeV$^{-2}$
which is a factor of 1.5 larger than the phenomenological Regge
slope. Possibly the discrepancy is eliminated when quantum
corrections to the transverse energy are taken into account (see
Sec.4).

It is interesting that our leading-order considerations do not
fix the transverse size of the chiral string $\sigma_0$ (see
Eq.(13)). In order to fix $\sigma_0$ we have to consider
corrections to the transverse energy (14).

First, there are corrections to the effective chiral action
itself. A popular way of modelling the higher derivatives terms
is to add to the kinetic energy term (4) the so-called Skyrme
term \cite{11},
\beq
S_{\rm Skyrme}=\ N_ce^2\int d^4x\sqrt{-g}\ g^{\alpha\beta}
g^{\mu\nu}\mbox{ Tr }[L_\alpha L_\mu][L_\beta L_\nu]\ ,
\eeq
where $e^2$ is a numerical constant of the order of unity.
Substituting the ansatz (10),(13) into (26) we get a correction
to $E_\perp$ of the form:
\beq
\delta E^{(1)}_\perp\ \sim\ \frac{e^2N_c}{\sigma^2_0}\ .
\eeq

Perhaps a more realistic four-derivative term of the effective
chiral action has been suggested in Ref. \cite{12}. It also leads
to a positive correction to the transverse energy of the same
form (27). Such a term prevents the chiral string from shrinking
to zero thickness.

What prevents it from infinite swelling in the transverse
direction? The answer to this question is less obvious.
Probably corrections in $\omega$ mentioned above are important
here. They come from taking into account the string end points
and large transverse distances $\sigma\ge1/\omega$. Since the
correction to the energy cannot depend on the direction of
rotation, at small $\omega$ it must be quadratic in $\omega$. On
dimension grounds one finds then that the correction should be
of the form:
\beq
\delta E^{(2)}_\perp\ \sim\ F^2_\pi\sigma^2_0\omega^2\ .
\eeq

Adding up (27) and (28) and minimizing their sum in $\sigma_0$
we obtain the transverse size
\beq
\sigma_0\ \sim\ \left(\frac{e^2N_c}{F^2_\pi\omega^2}
\right)^{1/4}\ \sim\ \frac1{F_\pi}\ (N_cJ)^{1/4}\ .
\eeq
This expression should be compared with that for the string
length
\beq
L=  \frac2\omega =  \frac{\sqrt{2J}}{F_\pi} = 4\sqrt{J\,\alpha'}.
\eeq

We see that the transverse size $\sigma_0$ of the rotating
chiral string grows with the angular momentum $J$ (thus
justifying the use of the long wave-length limit of the chiral
action) although more slowly than the longitudinal size $L$.

\section{Classical radiation by rotating chiral solitons}

Let us first of all show that  a strictly stationary rotating
soliton cannot exist: it has to lose its energy through
radiation of classical pion fields.

Indeed, let us consider the far-distance tail of the rotating soliton field
assuming that the pion field is already small there, so that one can
linearize the equation of motion. Then it is just the d'Alembert
equation. Assuming that the soliton rotates with the angular
velocity $\omega$ around the $z$ axis which means that the field
depends on $r$, $\vartheta$ and $\varphi'=\varphi-\omega t$, the
d'Alembert equation can be rewritten as
\beq
\Delta\,\pi^A-\omega^2\frac{\partial^2}{\partial\varphi'^2}\
\pi^A\ =\ 0\ .
\eeq

We look for its solution in the form
\beq
\pi^A(r,\vartheta,\varphi')\ =\ \sum_{\ell,m}R^A_{\ell m}(r)
Y_{\ell m}(\vartheta,\varphi')\ ,
\eeq
where radial functions satisfy the equation
$$
\left[\frac1{r^2}\ \frac d{dr}\ r^2\frac
d{dr}-\frac{\ell(\ell+1)}{r^2}+m^2\omega^2\right] R^A_{\ell
m}(r)\ =\ 0\ .
$$
Its solution are spherical Bessel functions $j_\ell(m\omega r)$
and $y_\ell(m\omega r)$ \cite{16}. Their asymptotics are
$j_\ell(z)\sim z^{-1}\sin(z-\ell\pi/2)$ and $y_\ell(z)\sim
-z^{-1}\cos(z-\ell\pi/2)$. Hence, the pion field (32) decreases
as $1/r$ at large distances. Note that such behavior is typical
for the radiation field. The energy and the angular momentum of
such field diverge at $r\to\infty$:
\begin{eqnarray}
E &=& \frac{F^2_\pi}2\int d^3r\left[(\partial_0\pi^A)^2 +
(\partial_i\pi^A)^2\right]\ =\ \infty\ ,\\
J_3 &=& -F^2_\pi\int d^3r\ \partial_0\pi^A\partial_\varphi\pi^A\
=\ \infty\ .
\end{eqnarray}

This is because we calculate here, in fact, not the energy and
the angular momentum of a soliton but of a soliton together with
its radiation field. The ``proper" energy and angular momentum of
the soliton itself can be found by subtracting the contribution
of the radiation to these quantities. Let us show how it is done
in the case of a spherically-symmetric soliton (a hedgehog). We
shall consider slow rotations, $\omega r_0\ll1$, where $r_0$ is
the typical size of the hedgehog. If $\omega r_0\ge1$ the
problem looses any sense: in this case the radiation is so
strong that one cannot speak of a stationary rotation. At $r\gg
r_0$ the pion field of the soliton is small, and one can use the
asymptotic form for the profile function of the hedgehog: $U=\exp
i(n\tau)P(r)$, $P(r)\to A/r^2$, $A=$const$\cdot r^2_0$. In the
range $r_0\ll r\ll1/\omega$ one has for the rotating soliton:
\begin{eqnarray}
&& \pi^1\ =\ \frac A{r^2}\sin\vartheta\cos(\varphi-\omega t),
\nonumber\\
&& \pi^2\ =\ \frac A{r^2}\sin\vartheta\sin(\varphi-\omega t),
\nonumber\\
&& \pi^3\ =\ \frac A{r^2}\cos\vartheta\ .
\end{eqnarray}

At the same time, the pion fields at $r\gg r_0$ must satisfy
the d'Alembert equation $\Box\pi^A=0$, to which the general
non-linear equation of motion is reduced when the fields are
linearized. In addition, at $r\to\infty$ the pion field must
satisfy the radiation condition,
\beq
\frac{\partial(\pi^A r)}{\partial r}+\frac{\partial(\pi^A r)}{\partial
t}\ =\ 0\ .
\eeq
In principle, one could as well look for a solution which has
the form of incoming waves at infinity. That would correspond to
a situation when one keeps the rotation by pumping energy from
infinity. But now we are interested in a free soliton which
looses energy and not {\it vice versa}. 

To find the field at large $r\geq 1/\omega$, one has to solve
the d'Alembert Eq.(31)
with the boundary conditions (35),(36) we get
to whioch the solution must reduce at $r\ll 1/\omega$. 
Such solution is readily found:
\begin{eqnarray}
&& \pi^1\ =\ \frac A{r^2}\sin\vartheta(\cos\alpha+\omega
r\sin\alpha), \nonumber\\
&& \pi^2\ =\ \frac A{r^2}\sin\vartheta(\sin\alpha-\omega
r\cos\alpha), \\
&& \pi^3\ =\ \frac A{r^2}\cos\vartheta\ , \qquad \alpha=\omega
r+\varphi-\omega t\ .  \nonumber
\end{eqnarray}
At $\omega r\ll 1$ it becomes the rotating soliton field (35)
and at $\omega r\gg 1$ the $\pi^{1,2}$ components fall of
as $1/r$ corresponding to the outgoing radiation waves. 

Let us find the intensity of the pion radiation by the soliton.
Using the general expression for the stress-energy tensor,
\beq
\vartheta_{\mu\nu}=\ \frac{F^2_\pi}2\left(\mbox{Tr }\partial_\mu
U\partial_\nu U^\dagger -\frac12\ g_{\mu\nu}\mbox{ Tr }\partial_\alpha
U\partial_\alpha U^\dagger\right)\ ,
\eeq
we get for the momentum flow at infinity:
\beq
\vartheta_{0i}\ \stackrel{r\to\infty}{\longrightarrow}\ F^2_\pi
\partial_0\pi^A\partial_i\pi^A\
 \to\ \frac{F^2_\pi A^2\omega^4\sin^2\vartheta\,
n_i}{r^2}\ .
\eeq
The momentum flow goes in the radial direction as 
it is proportional to the unit vector ${\bf n}$ but its intensity
depends on the angle $\vartheta$ from the rotation axis.

In  a slightly different context, this formula has been known to
Lorentz, coinciding in its angular and frequency dependence with
the intensity of the electromagnetic dipole radiation \cite{17}. It
is not accidental: owing to its Goldstone nature the
pion field couples to the isospin source through a gradient;
owing to gauge invariance only the field strength counts in the
e.m. radiation, which also has a gradient coupling to charges.
In the non-relativistic limit we are now considering only the
dipole component of the radiation survives in both cases.

The energy loss owing to the radiation is, according to the
energy-momentum conservation law $\partial^\mu
\vartheta_{\mu\nu}=0$,
\beq
-\frac{dE}{dt}=W=\lim\limits_{r\to\infty}\int d\Sigma_i\,
\vartheta_{0i}r^2=F^2_\pi A^2\omega^42\pi\int\limits^\pi_0
d\vartheta\sin^3\vartheta=\frac{8\pi F^2_\pi A^2\omega^4}3\ .
\eeq
We notice, that the coefficients $A$ in the asymptotics of
hedgehog profile function is directly related \cite{11} to the
nucleon axial constant $g_A=8\pi AF^2_\pi/3$ and the latter
quantity --- through the Goldberger--Treiman relation --- to the
pion-nucleon coupling constant $g_{\pi NN}=g_AM/F_\pi$. Using
these formulae, the radiation intensity can be rewritten as
\beq
W\ =\ \frac{3g^2_{\pi NN}\omega^4}{8\pi M^2}\ ,
\eeq
where $M$ is the mass of the soliton.

The radiation carries away not only the energy but also the
angular momentum which decreases as
\beq
\frac{dJ_3}{dt}=\lim\limits_{r\to\infty}\int d\Sigma_i\,
\varepsilon_{3jk}x_j\vartheta_{ik}=\ F^2_\pi\int d\Sigma_i
\frac{\partial\pi^A}{\partial x_i}\ \frac{\partial\pi^A}{
\partial\varphi}\ =\ - \frac W\omega\ .
\eeq

The ``proper" energy and angular momentum of a rotating soliton
can be found by subtracting volume integrals over, respectively,
the energy density and the angular momentum density of the
radiation \cite{17}:
\begin{eqnarray}
E^{\rm prop} &=& \int d^3r\,\vartheta^{prop}_{00}=\int d^3r
(\vartheta_{00}-n_i\vartheta_{i0})\ ,\\
J^{\rm prop}_3 &=& \int d^3r\,\varepsilon_{3jk}\vartheta^{prop}_{0j}
x_k=\int
d^3r\,\varepsilon_{3jk}(\vartheta_{oj}-n_i\vartheta_{ij})x_k\ .
\end{eqnarray}
It can be seen from Eqs.(37) that at $r\to\infty$ the integrands
in Eqs. (43,44) behave as
\begin{eqnarray}
\vartheta_{00}-n_i\vartheta_{i0} & \stackrel{r\to\infty}{=}&
\frac{F^2_\pi A^2}{r^4}\left(\frac6{r^2}+\omega^2\sin^2\vartheta
\right), \nonumber\\
\varepsilon_{3jk}(\vartheta_{0i}-n_i\vartheta_{ij})x_k &
\stackrel{r\to\infty}{=}& \frac{F^2_\pi A^2}{r^4}\ 2\omega\sin^2
\vartheta\ .
\end{eqnarray}
As a result both quantities, $E^{\rm prop}$ and $J_3^{\rm prop}$, 
are now
convergent. Moreover, it can be shown on general grounds that
these quantities satisfy familiar relations for slowly rotating
bodies:
\begin{eqnarray}
E^{\rm prop} &=& E^{\rm rest}+\frac{I\omega^2}2+O(\omega^4)\ ,
\nonumber\\
J^{\rm prop}_3 &=& I\omega+O(\omega^3)\ ,
\end{eqnarray}
where $E^{\rm rest}$ is the soliton rest mass and $I$ is its moment
of inertia. Eqs. (45) evidently correspond to the more general
relations (46) giving, in fact, the large-distance contributions
to $E^{\rm rest}$ and $I$. It is important that the energy loss
(40) is $O(\omega^4)$. To this accuracy rotation is
approximately stationary and $O(\omega^2)$ corrections to the
rest mass make sense.

Let us now calculate the lifetime of a highly excited $(J\gg1)$
rotational state of a hedgehog. This can be done in two
different ways: ``quantum" and ``classical". In the quantum
approach one has to calculate the transition amplitude between
the states $J$ and $J-1$ with the emission of a pion. We remind
the reader that the rotational excitations of a
spherically-symmetric soliton which we are now considering, have
isospin $T=J$ \cite{11}. The rotational wave functions are
Wigner $D$-functions which depend on  a unitary matrix $R$
characterizing the soliton orientation in spin-isospin space
\cite{11,12}:
\beq
\Psi^J_{J_3,T_3}(R)\ =\ \sqrt{2J+1}
(-1)^{J+J_3}D^J_{-T_3,J_3}(R)\ .
\eeq
The pion-soliton coupling is \cite{11,18}:
\beq
-\frac{3g_{\pi NN}}{2M}\ \frac12\mbox{ Tr }(R^+\tau^AR\sigma_i)\
ik_i\ ,
\eeq
where $k_i$ is the 3-momentum of the pion, $A$ is its isotopic
component. Sandwiching (48) between the initial and final wave
functions (47) we get for the $J\to J-1$ transition amplitude
squared (averaged over the initial and summed over the final
spin and isospin states)
\beq
\left(\frac{3g_{\pi NN}}{2M}\right)^2\ \frac{2J-1}{2J+1}\
\frac{|{\bf k}|^2}3\ .
\eeq
To get the decay width one has to multiply (49) by the phase
space factor $|{\bf k}|/2\pi$. We obtain
\begin{eqnarray}
\Gamma_{J\to J-1} &=& \frac{3g^2_{\pi NN}}{8\pi M^2}\
\frac{2J-1}{2J+1}\ \frac{M_{J-1}}{M_J}|{\bf k}|^3,\nonumber\\
|{\bf k}| &=&
\frac{\sqrt{(M^2_J-M^2_{J-1}+\mu^2_\pi)^2-4M^2_J\mu^2_\pi}}
{2M_J},\nonumber\\
M_J &=& M+\frac{J(J+1)}{2I}\ , \qquad M_J-M_{J-1}=\frac JI
=\omega\ .
\end{eqnarray}
In the chiral limit we put the pion mass $\mu_\pi=0$ and
obtain at $J\gg1$:
\beq
\Gamma\ =\ \frac{3g^2_{\pi NN}}{8\pi M^2}\ \left(\frac
JI\right)^3\ .
\eeq

Just as in the case of a highly excited state of an
atom, the lifetime can be also calculated from the classical
radiation theory. We have already found the energy loss per unit
time as due to the classical radiation of the pion field, see
Eqs.(40,41). The lifetime of a rotating soliton with given
$J\gg1$ can be determined as the time during which the energy of
the soliton decreases from $M_J$ to $M_{J-1}$. This prescription
is known as the Bohr correspondence principle: the lifetime of
an excited state of an atom is not the time during
which the electron looses all its energy through radiation but
only a small portion corresponding to the transition to the
nearest lower state. In our case we have therefore:
$$
\Gamma\ =\ \frac 1t\ =\ \frac W{M_J-M_{J-1}}\ =\ \frac{3g^2_{\pi
NN}}{8\pi M^2}\ \left(\frac JI\right)^3\ ,
$$
which coincides exactly with the quantum-mechanical result (51).
(In the last equation we have used Eq.(41) and the relation (46)
for the angular velocity, $\omega=J/I$). The same result follows
immediately from Eq.(42): one can determine the lifetime as the
time during which the soliton looses one unit of its angular
momentum:
$$
\Gamma\ = \ \left|\frac{dJ}{dt}\right| \frac1{J-(J-1)}\ =\ \frac
W\omega\ =\ \frac{3g^2_{\pi NN}}{8\pi M^2}\ \left(\frac
JI\right)^3\ . $$
We leave it for the reader to check that the time during which a
soliton looses one unit of isospin owing to the classical pion
radiation, is also given by the same formula.

We notice in passing that Eq.(50) gives the right numerical
value for the width of $\Delta$ resonance, $\Gamma_\Delta\simeq
110$ MeV. In this estimate we use experimental values of the $N$
and $\Delta$ masses and of the $\pi N$ coupling, $g_{\pi
NN}\simeq13.6$. For the exotic $J=T=5/2$ state with
$M_{5/2}\simeq1700$ MeV Eq.(50) predicts $\Gamma(5/2\to3/2)
\simeq760$ MeV. Such a big width explains perhaps why the
$(5/2,5/2)$ resonance has not been definitely observed.

Let us summarize. We start with a spherically-symmetric chiral
soliton (a skyrmion) which presumably reproduces the nucleon. We
try to rotate it. If $\omega$ is small, the form of the soliton
does not change. It remains spherically-symmetric, hence, only
$T=J$ states are allowed. However a rotating soliton inevitably
radiates pion fields. We have demonstrated that the classical
radiation is in direct correspondence with the
quantum-mechanical calculation of the widths.

At $\omega r_0\sim1$ (where $r_0$ is the characteristic size of
the soliton) the radiation becomes very strong, the widths blow
up. Fast rotating hedgehogs, actually with $T=J=O(N_c)$, do not
exist. Deformation due to centrifugal forces cannot be further
ignored. If we insist on getting a soliton with still larger
$J$, it must be stretched in the direction perpendicular to the
rotation axis. But when we pass to a cigar-like (eventually to a
string-like) soliton, the angular velocity $\omega$ is no more
proportional to $J$. On the contrary, it decreases with $J$, see
Eq.(23). Since the radiation grows with $\omega$, it means that
a rotating string-like soliton may be relatively stable with
respect to classical radiation (we shall estimate the
corresponding width in a moment). A non-spherically symmetric
soliton does not necessarily have isospin $T=J$. In fact $T$ may
remain fixed. We come to linear Regge trajectories described in
Sec. 2.

If we plot angular velocity $\omega$ as a function of $J$ we see
that there exists a maximal $\omega$ at which a crossover from a
spherically-symmetric to a string-like soliton occurs, see
Fig. 5.

Finally, let us estimate the intensity of the classical
radiation of a rotating string-like soliton. Similar to the case
of a spherically-symmetric hedgehog discussed above, one has to
find the solution of the d'Alembert equation together with the
radiation condition (36) and with the boundary conditions being
the soliton field itself. Since the soliton field decreases far
away from the string axis as $\sigma_0/\sigma$ (see Eq.(13))
where $\sigma$ is the distance from the axis, the radiation
intensity $W$ is proportional to $\sigma^2_0$ where $\sigma_0$
is the string thickness. On dimensional grounds we find then:
\beq
W\ \sim\ F^2_\pi\sigma^2_0\omega^2\ \sim\ F^4_\pi\sigma^2_0/J\ .
\eeq
The width due to classical radiation is
$\Gamma=W/(M_J-M_{J-1})$. Taking into account that
$M^2_J=J/\alpha'$ and that $1/\alpha'\sim F^2_\pi$ we
get:
\beq
\Gamma^{\rm class}\ \sim\ \frac{F^3_\pi\sigma^2_0}{\sqrt J}\ .
\eeq
Finally, we use Eq.(29) for the transverse size $\sigma_0$ and
find for the width of a state $J$:
\beq
\Gamma^{\rm class}\ \sim\ F_\pi\ \sqrt{N_c}\ ,
\eeq
which is linear in $N_c$ but independent of $J$ (cf. with the
width of a spherically-symmetric rotator (Eq.(51)), which grows
as $J^3$). The width-to-mass ratio is
\beq
\frac{\Gamma^{class}}M\ \sim\ \sqrt{\frac{N_c}J}\ .
\eeq

Let us mention that there is also a purely quantum contribution
to the width of a Regge resonance which we calculate in the next
section.

\section{Quantum corrections, zero modes and all that}

In Sec. 2 we have found an extremum of the chiral action (4),
corresponding to a thick rotating string. The solution is given
by Eqs. (19,13). Let us expand the chiral field $U$ near the
classical solution which we denote by $U_0$. Introducing
hermitean quantum fluctuations $r$:
\beq
U\ =\ U_0\left(1+ir-\frac{r^2}{2}+\cdots\right)\ ,
\eeq
we obtain for the action
\beq
S=-\frac{2\pi^2F^2_\pi}\omega\int\! dt\;+\;\frac{F^2_\pi}4\int
d^4x'g^{\mu\nu}(x')\mbox{ Tr}\left\{\partial_\mu r\partial_\nu
r+i\partial_\mu r[r,L_\nu]+O(r^3)\right\}\ .
\eeq
Neglecting terms of the order of $\omega y'$ and $\omega z'$ in the
metric tensor given by Eq.(3) (to that accuracy the classical
equation of motion has been solved in Sec. 2) we get the following
quadratic form for quantum fluctuations:
\begin{eqnarray}
S^{(2)}&=& \frac{F^2_\pi}4\int d^4x'\mbox{ Tr}\bigg\{
(\partial_tr)^2-\frac{2\omega x'}{\sqrt{1-\omega^2x'^2}}\
\partial_tr\left(\partial_{\tilde y}r+i[r,L_{\tilde y}]\right)
\nonumber\\
&-& (\partial_{x'}r)^2-\partial_{\tilde
y}r\left(\partial_{\tilde y}+i[r,L_{\tilde y}]\right)
-\partial_{\tilde z}r\left(\partial_{\tilde z}r +i[r,L_{\tilde
z}]\right)\bigg\}\ .
\end{eqnarray}

We remind the reader that the primes refer to the body-fixed
frame and the tildes to the Lorentz-contracted coordinates 
(see Sec. 2).

Let us consider the transverse Laplace operator
\beq
\Delta_\perp\ \equiv\ \frac{\partial^2}{\partial\tilde y^2}
+\frac{\partial^2}{\partial\tilde z^2}-i\left[L_{\tilde y},
\frac\partial{\partial\tilde y}\cdots\right]-i\left[L_{\tilde z}
,\frac\partial{\partial\tilde z}\cdots\right]\ .
\eeq
Let its eigenfunctions be $v^A_n\tau^A$ with eigenvalues
$\lambda_n$:
\beq
-\Delta_\perp v^A_n\tau^A\ =\ \lambda_nv_n^A\tau^A\ .
\eeq
The functions $v_n^A(\tilde y,\tilde z)$ form a complete set of
ortho-normalized functions. Therefore, we can decompose 
the general fluctuation field $r^A$ as
\beq
r^A(t,x',y',z')\ =\ \sum_n c_n(t,x')v^A_n(\tilde y,\tilde z)\ .
\eeq
Substituting (61) into Eq.(58) we get:
\beq
S^{(2)}\ =\ \frac{F^2_\pi}2\int dtdx'\sqrt{1-\omega^2x'^2}
\sum_n\left[\left(\frac{\partial c_n}{\partial t}\right)^2
-\left(\frac{\partial c_n}{\partial x'}\right)^2 -\lambda_nc^2_n
\right]\ .
\eeq
Note that the second term in Eq.(58) cancels out. The
relativistic square root arises here when one passes from $y'$
to $\tilde y$.

Eq.(62) is the action for excitations which ``live" on the string
and have masses $\sqrt{\lambda_n}$. Their zero-point
oscillations give the quantum correction to the transverse
energy (or energy per unit length) of the string:
\beq
E^{\rm quant}_\perp\ =\ \sum_n\int\frac{dk_x}{2\pi}\ \frac12\left(
\sqrt{\lambda_n+k^2_x}-\sqrt{\lambda_n^{(0)}+k^2_x}\right)\ ,
\eeq
where $\lambda^{(0)}_n$ are eigenvalues of the free Laplace
operator.

If there is a negative eigenvalue $\lambda_-$, it
leads to a non-zero imaginary part of the energy. It means that
the soliton is unstable with respect to quantum fluctuations in
a particular direction in Hilbert space. The
corresponding width is
\beq
\Gamma\ =\ 2\mbox{ Im }E\ =\ 2\int dx'\sqrt{1-\omega^2x'^2}
\int\limits^{\sqrt{|\lambda_-|}}_{-\sqrt{|\lambda_-|}} \frac{dk_x}{2\pi}\
\frac12\sqrt{|\lambda_-|-k^2_x}\ =\ \frac{\pi|\lambda_-|}{8\omega}\ ,
\eeq
where $\lambda_-$ is the negative eigenvalue of the transverse 
Laplace operator (59).

Let us give a qualitative argument that there must be at least
one negative eigenvalue. The point is that, while in 3
dimensions a hedgehog belongs to a non-trivial homotopy class
thanks to $\pi_3(SU_2)=Z$, in 2 dimensions it is reducible by a
continuous deformation to a zero field, since $\pi_2(SU_2)=0$.
Therefore, it is natural to expect that the extremum we have
found in Sec. 2 is in fact a saddle point. The exact calculation
below confirms this expectation.

In order to perform exact calculations one has to diagonalize
the $\Delta_\perp$ operator (59). If the external field $U_0$ is
a 2-dimensional hedgehog given by Eqs.(10,13) $\Delta_\perp$
commutes with $K=(T+L)_{x'}$ --- the projection of isospin plus
orbital moment on the string axis. Let us parametrize the
transverse coordinates: $\tilde y=\sigma\cos\Psi$, $\tilde
z=\sigma\sin\Psi$. Using the concrete transverse profile
function $P(\sigma)=2\arctan(\sigma_0/\sigma)$ and putting
temporarily $\sigma_0=1$ we obtain:
\begin{eqnarray}
\Delta_\perp &=& \left( \frac{\partial^2}{\partial\sigma^2}+
\frac1\partial\frac\partial{\partial\sigma}+\frac1{\sigma^2}
\frac{\partial^2}{\partial\Psi^2}\right)-\frac i{\sigma^2+1}
\left[e^{i\Psi}\tau^- +e^{-i\Psi}\tau^+, \frac\partial{\partial
\sigma}\cdots\right] \nonumber\\
&+& \frac1{\sigma(\sigma^2+1)}\left[4\sigma r^1+i(\sigma^2-1)
(e^{i\Psi}\tau^- -e^{-i\Psi}\tau^+),\frac\partial{\partial\Psi}
\cdots\right]\ ,
\end{eqnarray}
where $\tau^{\pm}=\tau^2\pm i\tau^3$. Apparently this operator
does not mix states with different $K$. Therefore, we look for
the eigenfunctions in the form:
\beq
v^A_K(\tilde y,\tilde z)\ =\ if_K(\sigma)e^{iK\Psi}\tau^1
+g_K(\sigma)e^{i(K+1)\Psi} \tau^-
+h_K(\sigma)e^{i(K-1)\Psi} \tau^+\ .
\eeq

The eigenvalue Eq. (60) becomes a system of 3 ordinary
equations:
\begin{eqnarray}
&&  -f''_K-\frac1\sigma f'_K
+\frac{K^2}{\sigma^2}f_K +\frac4{\sigma^2+1}(g_K-h_K)'
\nonumber\\
&& \quad -\quad  \frac{4(\sigma^2-1)}{\sigma
(\sigma^2+1)^2}[(K-1)h_K +(K+1)g_K]\ =\ \lambda f_K; \\
&& \hspace{-1cm}
-g''_K-\frac1\sigma g'_K +\frac{(K+1)^2}{\sigma^2}g_K -
\frac2{\sigma^2+1}f'_K -\frac{8(K+1)}{(\sigma^2+1)^2}g_K
-\frac{2K(\sigma^2-1)}{\sigma(\sigma^2+1)^2}f_K\ =\ \lambda g_K;
\nonumber\\
&& \hspace{-1cm}
-h''_K-\frac1\sigma h'_K+\frac{(K-1)^2}{\sigma^2}h_K
+\frac2{\sigma^2+1}f'_K +\frac{8(K-1)}{(\sigma^2+1)^2}h_K
-\frac{2K(\sigma^2-1)}{\sigma(\sigma^2+1)}f_K\ =\ \lambda h_K.
\nonumber \end{eqnarray}
It can be seen that at $K=0$ this system splits into an equation
for $(g+h)_0$ and a system of two equations for $f_0$ and $(g-h)_0$:
\begin{eqnarray}
-f''_0-\frac1\sigma f'_0+\frac4{\sigma^2+1}(g-h)'_0
-\frac{4(\sigma^2-1)}{\sigma(\sigma^2+1)^2}(g-h)_0 &=& \lambda
f_0 \\
-(g-h)''_0-\frac1\sigma(g-h)'_0 +\frac1{\sigma^2}(g-h)_0
-\frac4{\sigma^2+1}f'_0 -\frac8{(\sigma^2+1)^2}(g-h)_0 &=&
\lambda(g-h)_0 \nonumber
\end{eqnarray}
and
\beq
-(g+h)''_0-\frac1\sigma (g+h)'_0 +\frac1{\sigma^2}(g+h)_0
-\frac8{(\sigma^2+1)^2}(g+h)_0\ =\ \lambda(g+h)_0\ .
\eeq

It is not clear if these equations can be solved analytically.  
However, the zero modes $(\lambda=0)$ can be found 
exactly from symmetry considerations.

\subsection*{Zero modes}

We expect 6 zero modes of the operator $\Delta_\perp$: two
translations, three global isospin rotations and one dilatation.
The last one is related to the fact that the transverse energy
does not depend on the string thickness $\sigma_0$.

The infinitesimal variation of the chiral field with translation
of the string axis in the $(\tilde y,\tilde z)$ plane is
$$ \delta U\ =\ -\frac{\partial U_0}{\partial x_\alpha}\ \delta
x_\alpha\ \equiv\ U_0ir_\alpha\delta x_\alpha\ , \qquad
x_\alpha=\ (\tilde y, \tilde z)\ .  $$
It corresponds to two translational zero modes $(\alpha=1,2)$:
\beq
r_\alpha\ =\ iU^+_0 \partial_\alpha U_0\ =\ L_\alpha\ .
 \eeq
One can choose their combinations $r_{\pm}=i(L_{\tilde y}\pm
iL_{\tilde z})/2$. Being written in the form (66) these
functions have quantum numbers $K=\pm1$, respectively. 
The corresponding $f,g,h$ functions are:
\begin{eqnarray}
f_{+1} &=& \frac{-2\sigma}{(\sigma^2+1)^2}\ , \quad g_{+1} =\
\frac{\sigma^2}{(\sigma^2+1)^2}\ , \quad h_{+1}=\
\frac1{(\sigma^2+1)^2}; \nonumber\\
f_{-1} &=& \frac{2\sigma}{(\sigma^2+1)^2}\ , \quad g_{-1}=\
\frac1{(\sigma^2+1)^2}\ , \quad h_{-1}=\
\frac{\sigma^2}{(\sigma^2+1)^2}\ .
\end{eqnarray}
These functions satisfy Eqs. (67) with $\lambda=0$, hence they
are, indeed, zero modes.

Similarly, there are three rotational zero modes,
\beq
r^1=\ U^+_0\tau'U_0-\tau'\ ,\qquad r^{\pm}=\ U^+_0\tau^{\pm}U_0
-\tau^{\pm}\ .
\eeq
Being standartized to the form given by Eq.(66) these functions
appear to have quantum numbers $K=0$ and $K=\pm1$, respectively.
We have:
\begin{eqnarray}
&& f_0 =\ \frac{4\sigma^2}{(\sigma^2+1)^2}\ , \quad (g-h)_0=\
-\frac{\sigma(\sigma^2-1)}{(\sigma^2+1)^2}\ ,\quad (g+h)_0=0\ ;\\
&& f_{\pm1}=\ \mp\frac{\sigma(\sigma^2-1)}{(\sigma^2+1)^2}\ ,
\quad g_{\pm1}=\ \pm\frac{\sigma^2}{(\sigma^2+1)^2}\ , \quad
h_{\pm1}=\ \mp\frac{\sigma^2}{(\sigma^2+1)^2}\ .
\end{eqnarray}
The functions (73) satisfy Eqs. (68) with $\lambda=0$ and the
function (74) satisfy Eqs.(67) at $K=\pm1$ also with a zero
right-hand side.

Finally, the dilatational zero mode is
\beq
r\ =\ -iU^+_0\ \frac{\partial U_0}{\partial\sigma_0}
\bigg|_{\sigma_0=1}\ =\ \frac{2(n\tau)\sigma}{\sigma^2+1} \ ,
\eeq
which corresponds to a state with $K=0$:
\beq
f_0=\ 0\ , \quad (g-h)_0=\ 0\ , \quad (g+h)_0=\
\frac\sigma{\sigma^2+1} \ .
\eeq
This function is a zero mode of Eq.(69).

We notice that the rotational (Eqs.(73,74)) and the
dilatational (Eq.(76)) zero modes, in contrast to the
translational ones, decrease as $1/\sigma$ at large distances
from the string axis. Therefore, they are not normalizable and
cannot, strictly speaking, be considered as zero modes.
Actually, they belong to the continuous spectrum.

For this reason, connected with the particular form of our
``instanton" solution (10), one is left only with the translation
zero modes, and the low-energy effective string theory is just
that of Nambu (Eq.(16)). In practical terms it means, in
particular, a degeneracy of trajectories in isospin --- a
property which seems to be realized in nature but looks totally
unexpected for a string made of chiral fields!

It should be stressed that since the resulting string theory is
an effective (not microscopic) one,  taking into account only
the long-wave excitations of the string, one should not be
confused by ghosts, tachyons and other inconsistencies which 
arise in the Nambu string at $d=4$.

\subsection*{Negative mode}

As anticipated above from topological considerations, the
operator $-\Delta_\perp$ has a negative eigenvalue. It belongs
to the $K=0$ sector (Eq.(68)). There are no negative eigenvalues
in other sectors. Solving Eq.(68) numerically we find
$\lambda_-\simeq-2.64$ (in units of $1/\sigma^2_0$). According to
Eq.(64) we get for the quantum width:
\beq
\Gamma\,^{\rm quant}\ =\ \frac{2.64\pi}{8\sigma^2_0\omega}\ =\
\frac{2.64\ \sqrt J}{8\sqrt2 F_\pi\sigma^2_0}\ ,
\eeq
where the relation between $\omega$ and $J$ (Eq.(23)) has been
used. Further on, if we take a slowly growing transverse size
$\sigma_0$ as given by Eq.(29) we get:
\beq
\Gamma\,^{\rm quant}\ \sim\ \frac{F_\pi}{\sqrt{N_c}}\ ,
\eeq
which is independent of $J$ and is $N_c$ times smaller than the
classical width $\Gamma^{\rm class}$, see Eq.(54). It means
that the widths of the resonances lying on Regge trajectories
are mainly determined by the classical radiation of pion fields.

\section{Discussion and conclusions}

In the last years there has been progress in understanding
nucleons as chiral solitons. The $\Delta$ resonance with
$T=J=3/2$ fits nicely the idea of being the first
rotational excitation of the nucleon soliton \cite{11,12}.
Theoretically speaking, rotations can be considered as quantum
corrections as long as $J\ll N_c$.

In Sec. 3 we have considered the range $1\ll J\ll N_c$. Alhough
hardly realized in nature, this range of angular momenta is very
interesting from the theory point of view since the rotation can be 
considered simultaneously in quantum and classical framework. 
The main result of Sec. 3 is that rotating chiral solitons inevitably
radiate pion fields quite similarly to the rotating electric
charges. We have calculated explicitly the radiation intensity
and hence the ``classical" width of a state with given $J$. In
accordance with Bohr correspondence principle the widths
calculated classically coincide at $J\gg1$ with the exact
quantum-mechanical result.

At $J\sim N_c$ the rotation of a spherical-symmetric soliton
becomes so fast that the classical radiation of pions  blows up 
and the widths become comparable to the masses. In order to 
``survive", the soliton with $J\geq N_c$ has to expand in a direction 
perpendicular to the rotation axis. This is exactly what one expects from 
the action of centrifugal forces. The angular velocity of an expandable 
cigar-like soliton {\em decreases} with the growth of $J$. As a result the 
lifetime of a rotating soliton becomes stable with $J$ even as one goes
to higher and higher values of angular momentum. 

Unfortunately, the crossover region at $J\sim N_c$ is too
complicated to be studied analytically. A significant
simplification is achieved at $J\gg N_c$. In this case a
string-like analytical solution of the equation of motion for a
rotating chiral field can be found (Sec.~2). Actually the
solution looks more like a double-blade kayak paddle, being
squeezed at the end points by Lorentz contraction. Knowing the
transverse pion field distribution inside the string (or the
kayak paddle) one can easily calculate the energy per unit
length or the ``string tension" and hence the Regge slope. We
find $\alpha'\simeq(8\pi^2F^2_\pi)^{-1}\simeq1.45$ GeV$^{-2}$
which is a factor of 1.5 larger than the phenomenological value
$\alpha'=0.8-0.9$ GeV$^{-2}$. It means that our resonances
are $\sqrt{1.5}$ times lighter, for given $J$, than in reality. 
The discrepancy is possibly eliminated when quantum corrections 
to the classical soliton energy are added. A systematic 
way to study quantum fluctuations is outlined in Sec. 4.

We do not also exclude the possibility that other, perhaps more
adequate classical solutions may be found. For example, an
obvious modification of our ansatz (10) is to take the
winding number $n_w$ in the instant transverse plane to the
string axis to be larger than 1 \cite{19}. One gets then for the
transverse profile function
$P(\sigma)=\arctan(\sigma_0/\sigma)^{n_w}$ and for  the Regge
slope $\alpha'=(8\pi^2F^2_\pi n_w)^{-1}$. [Curiously, the
phenomenological value of $\alpha'$ corresponds to 
$n_w=\frac{3}{2}$.] 
Ultimately, the true
classical solution should be chosen from the requirement of
minimal width due to classical radiation -- the
minimum-of-mass criterium is senseless for the whole Regge
trajectory. For a given classical rotating solution,
one has to investigate the quadratic form for
quantum oscillations about it and extract
the zero modes since they determine the low-energy string
theory (see Sec. 4). Finally, one should check that the
quantization of the effective string theory gives the right quantum
numbers. 

In this paper we did not specify what Regge trajectories we were
dealing with -- meson or baryon. At $1\sim J\ll N_c$ only
baryons can be treated as chiral solitons since there is no
large parameter in the meson case allowing for a semiclassical
approach. However at $J\ge N_c$ both baryon and meson resonances
can be understood as rotating solitons discussed in this paper,
with quantum corrections suppressed as $1/N_c$. 

Where does the difference between meson and baryon trajectories
come in? To answer this question we have to go one step back 
and consider the$\sigma$-model-type quark--pion lagrangian 
suggested in Ref. \cite{12}. Integrating out quarks one gets the effective
chiral action whose derivative expansion starts with the kinetic
energy term (1).

In terms of the quark-pion lagrangian the problem of getting a
classical extremum is to find the pion field providing an
extremum for the quark determinant in that field. If we take the
pion field in the form of a rotating string, the problem is
reduced to solving the two-dimensional transverse Dirac equation
in the background chiral field given by Eq.(10). Thanks to 
two dimensions there is always a discrete bound state whose position 
$\varepsilon_0$ depends on the string thickness $\sigma_0$. 
All levels are $N_c$ times degenerate. The squared eigenvalues of the 
transverse Dirac operator $\varepsilon^2_n$ play the r\^ole of 
the squared masses for fermions which ``live" on the string 
(cf. the boson excitations considered in Sec. 4). Their energies are
$\pm\sqrt{\varepsilon^2_n+k^2_x}$.

The zero baryon number states (mesons)  are obtained by filling 
in all energy levels which come from the negative-energy 
continuum when the trial pion field is switched in (see Fig. 6). 
It can be easily shown that at $\sigma_0M\gg1$ where $M$ is the 
dynamical quark mass, $M\simeq 350$ MeV \cite{12}, the 
aggregate energy of the occupied levels coincides with the 
kinetic-energy term of the chiral action, given by 
Eq.(24). Therefore, nothing has to be modified in the above.

To get the unity baryon number states (baryons)  one has to fill in the 
lowest level coming from the upper Dirac continuum (also shown in
Fig. 6). It increases the energy given by Eq.(24), {\it viz.} $E=O(J)$, 
by $|\varepsilon_0|$. However this quantity is independent of $J$
and hence does not alter the Regge slope. 

We conclude that in our chiral approach meson and baryon
Regge trajectories automatically have the same slope.

\vskip 0.5true cm

To summarize: We have shown how a rotating relativistic 
string can be formed from massless chiral fields, giving rise to
linear Regge trajectories. Both the longitudinal and transverse
sizes of the string grow with angular momentum. Therefore,
large-J resonances are huge in all directions, and it seems
reasonable to understand and describe them in terms of
the lightest degrees of freedom -- the pions -- and not as gluonic
strings which anyhow have to break because of the existence of
light pions. 

\vskip 0.5true cm

We are grateful to M. Bander, A.E. Kudrjavtsev, 
V.A. Kudryavtsev, P.V. Pobylitsa and especially to M.I. Eides for 
useful discussions.

\newpage
\begin{figure}[t]
\epsfig{file=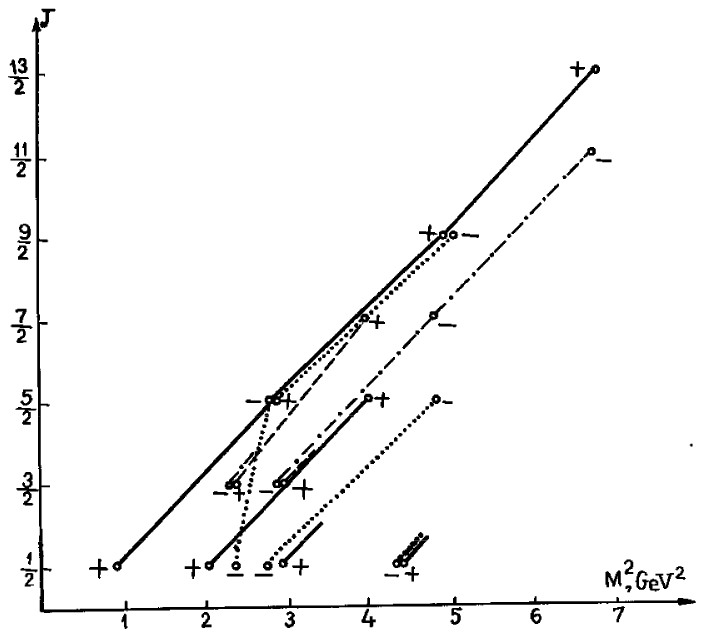,width=9.5cm}
\vspace{-.01cm}\caption{
Baryon states with isospin $T=1/2$. The signs
denote parity;  their sizes correspond to the status of a
resonance.}
\end{figure}
\begin{figure}[t]
\epsfig{file=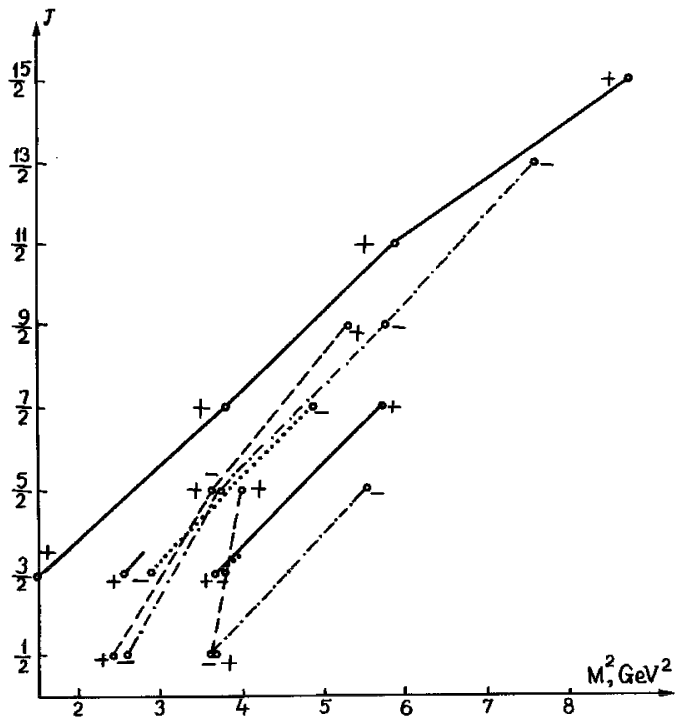,width=9.5cm}
\vspace{-.01cm}\caption{
Baryon states with isospin $T=3/2$. The signs
denote parity;  their sizes correspond to the status of a
resonance.}
\end{figure}

\begin{figure}[t]
\epsfig{file=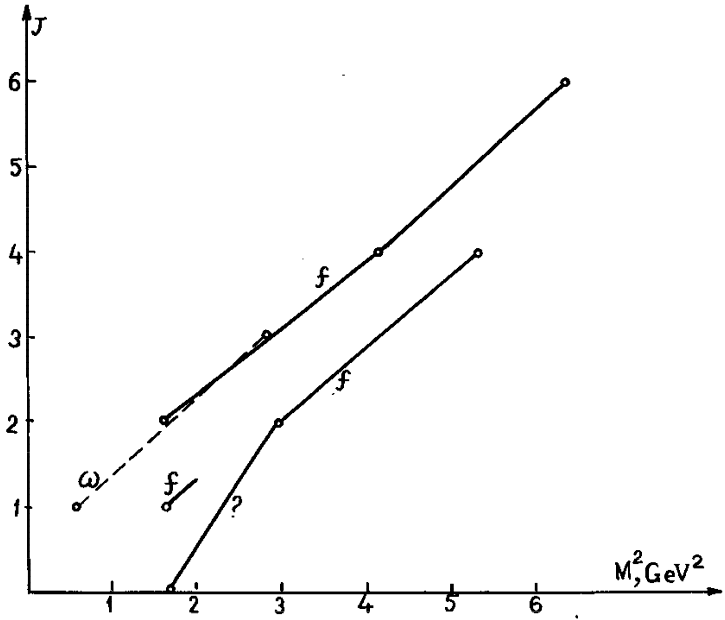,width=10cm}
\vspace{-.01cm}\caption{Meson states with $T=0$. $f$-trajectory has
$T^G=0^+$; $\omega$-trajectory has $T^G=0^-$. Both have
``natural" parity. Omitted are states with presumably large
admixtures of strange quarks or having a low status.}
\end{figure}

\begin{figure}[t]
\epsfig{file=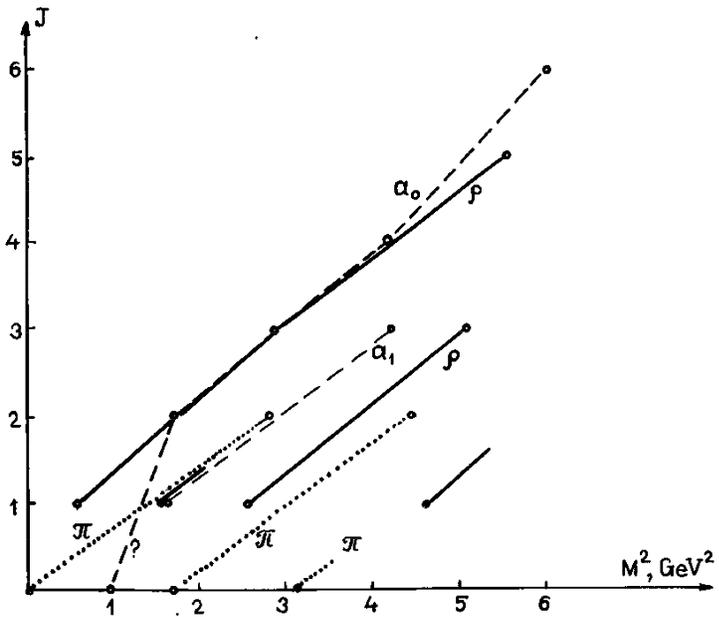,width=10cm}
\vspace{-.01cm}\caption{Meson states with $T=1$. $\rho$-trajectory has
$T^G=1^+$, $\pi$- and $a_0$-trajectories have ``natural" whereas
$\pi$- and $a_1$-trajectories have ``unnatural" parity.}
\end{figure}

\begin{figure}[t]
\epsfig{file=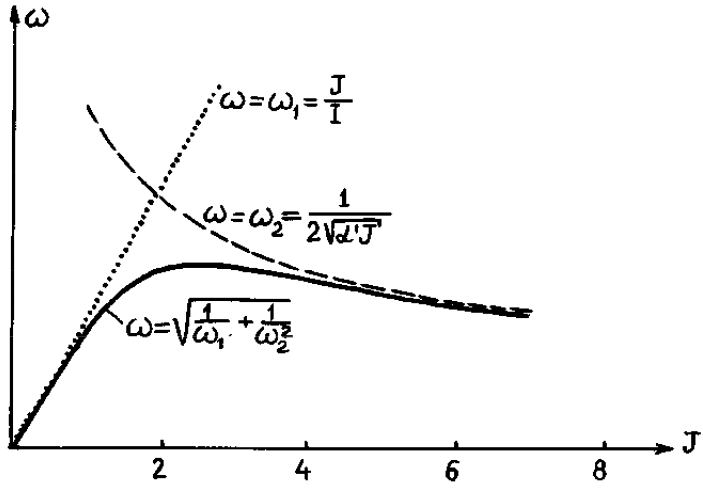,width=10cm}
\vspace{-.01cm}\caption{Dependence of the angular velocity of a rotating
soliton on its momentum. $\omega_{1,2}$ are two regimes
corresponding to low and large $J$, respectively. The solid line
is a possible interpolation.}
\end{figure}

\begin{figure}[t]
\epsfig{file=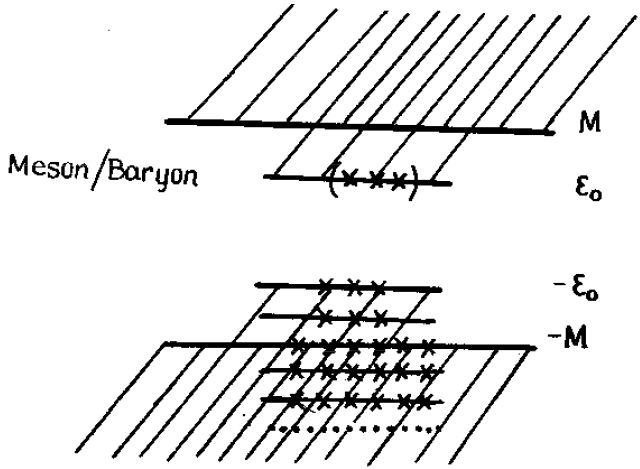,width=10cm}
\vspace{-.01cm}\caption{Quark spectrum in a background string-like 
chiral field. The upper and lower Dirac continua are shaded. 
The crosses denote occupied levels, $N_c$ times degenerate in 
color. The upper level is filled in baryons and empty in mesons. 
It results in a shift of baryons massses as compared to mesons, 
by a value independent of the angular momentum.}
\end{figure}

\end{document}